\documentclass[twocolumn]{emulateapj}

\usepackage{amsmath}
\usepackage{graphicx}
\usepackage{float}
\usepackage{booktabs}
\usepackage{epstopdf}
\usepackage{mathtools}
\usepackage{placeins}

\newcommand{\rmclass}{{\rm class}}
\newcommand{\rmcost}{{\rm Cost}}

\defcitealias{isanun2014fats}{{FATS}: {Feature Analysis for Time Series}}

\begin{document}

\title{Meta classification for variable stars}
\author{Karim Pichara$^{1, 3, 2}$, Pavlos Protopapas$^{2}$ and Daniel Le\'on$^{1}$}
\affil{\altaffilmark{1}Computer Science Department, Pontificia Universidad Cat\'olica de Chile, Santiago, Chile}
\affil{\altaffilmark{2}Institute for Applied Computational Science, Harvard University, Cambridge, MA, USA}
\affil{\altaffilmark{3}Millennium Institute of Astrophysics, Chile}

\begin{abstract}
The need for the development of automatic tools to explore astronomical databases has been recognized since the inception of CCDs and modern computers. Astronomers already have developed solutions to tackle several science problems, such as automatic classification of stellar objects, outlier detection, and globular clusters identification, among others. New science problems emerge and it is critical to be able to re-use the models learned before, without rebuilding everything from the beginning when the science problem changes. In this paper, we propose a new meta-model that automatically integrates existing classification models of variable stars.
The proposed meta-model incorporates existing models that are trained in a different context, answering different questions and using different representations of data. Conventional mixture of experts algorithms in machine learning literature can not be used since each expert (model) uses different inputs. We also consider computational complexity of the model by using the most expensive models only when it is necessary. 
We test our model with EROS-2 and MACHO datasets, and we show that we solve most of the  classification challenges only by training a meta-model to learn how to integrate the previous experts.
\end{abstract}


\section{Introduction}

The scientific community is dealing with massive amounts of digital information and astronomy is not an exception (see for example \citep{Cook1995, 2002A&A...389..149D, Kaiser2004SPIE, Ivezic2008, Udalski2008AcA}). 
It is practically impossible to analyze the vast amount of data, generated by modern telescopes and surveys, without the help of machines. This is done either with the use of simple algorithmic solutions or machine learning approaches.  A particular example of such automatic methods is the automatic classification of variable objects \citep{Richards:2011, Bloom:2011, Bloom2:2011, Kim:2012, Pichara_QSO:2012, Pichara_Miss:2013}, which is the focus of this paper. Automatic classification of variable stars makes it possible to speed scientific discoveries through an initial labelling, thus allowing astronomers to have a selection of lightcurves of interest for further study and analysis. 
Many solutions in automatic classification have been proposed \citep{Richards:2011, Long2012, Pichara_QSO:2012, Butler2011AJ, Kim2014}. These models, called  {\em experts} in this paper, classify to a  subset of possible classes, using a set of specific variables (hereinafter called features) that represent the lightcurves. In this work, we aim that future models can take advantage of those models in solving new challenges. 
As an example, suppose that we have the following models:
\begin{itemize} 
\item A model that classifies objects in quasars and no-quasars, trained with a specific set of features \citep{Kelly:2009, Kim:2012, Pichara_QSO:2012}. 
\item A model that separates  periodic from non-periodic objects \citep{Huijse2012, Protopapas2015, Kim2014}. 
\item A general purpose classifier that can classify  (with a bit lower accuracy)  many different variability classes \citep{Richards:2011, Long2012, Pichara_Miss:2013}.
\item A model that classifies RR Lyrae \citep{Gran2015} from the rest. 
\item A model that identifies microlensing and eclipsing binaries \citep{Belokurov2003}. 

\end{itemize}
 Assuming we need to create a model that classifies RR Lyrae, Eclipsing Binaries, Be stars, and quasars, it is apparent that there is a lot of intersection between the new desired model and the previous models we have. Therefore, we should be able to solve our new challenge without the need to build a totally new model.

 The idea of mixing many different models is very old in machine learning literature \citep{Jordan1993, Rasmussen1991, Meir1996, Breiman:2001, Kuncheva2007, Kuncheva_Whitaker, Bishop2012, Chamroukhi2015}. These approaches are guided by the  `divide and conquer' principle, in which each expert focuses on a particular area of feature space. Most of the solutions proposed from machine learning literature assume the same context for each of the experts. In other words, they deal with data represented in the same feature space (same variables describing the data), and in most of the cases with the same set of predicted classes. To the best of our knowledge, there is not a mixture of experts solution that combines all the context variants we mention before, together with the efficient management of the computational complexity of the experts.

  Our approach is based on a very simple idea of empirically estimate how fit a model is in a given scenario. In other words, the best we can do in learning how to combine different experts is to try them in different cases and evaluate their results. This idea allows us to avoid the need for the understanding of the internal structure of the experts, which can be very costly. Our approach first creates a meta-dataset containing all the experts' outputs obtained from the initial training dataset. We search for patterns in the classification results and we model these patterns to predict the lightcurve classes. This is the meta-classifier, which integrates the outputs from the experts to make a final prediction.

The idea of studying models outputs was used before, but in different contexts, such as anomaly detection \citep{Nun2014} and measurements of diversity \citep{Kuncheva_Whitaker}. 

Furthermore, the integration model has to be easily understood and interpretable. It is not desirable to have a `black box'  that integrates the decision in an unknown and confusing way because we can not gain intuition on how the model is deciding, and how each expert is contributing to the final decision. Decision trees are very suitable for simple decision patterns; each node represents a question, and each directed edge pointing out from a node represents an answer from the node. There are many algorithms proposed to train a decision tree \citep{Quinlan:1986, Quinlan:1993}, but those algorithms just focus on optimising the classification accuracy rather than consider the cost of each question done on each of the nodes. 

    There are many known techniques that aim to describe lightcurves as vectors of real numbers (features) by trying to extract the maximum information from lightcurves while maximizing the classification performance. In a recent work, \cite{Nun2015} presented an automatic tool that calculates more than 60 such features. Depending on the classification task, some features are more useful than others, and some features are more computationally costly than others. One example of computationally costly features are the coefficients for a continuous autoregressive model  \citep{Kelly:2009, Pichara_QSO:2012}. However, these features have been shown to be helpful in differentiating quasars from other stars. Another example is the correntropy kernelized periodogram \citep{Huijse2012}, which has been used  to classify periodicity classes. Both kinds of features are more expensive than others, but both present significant improvements in the classification tasks.
      
  In our mixture of experts set up, each of the models solves different problems using different features with different estimation cost. The cost of classification from the mixture of experts can be calculated. Every time we ask a model to classify a given lightcurve, we know the features used by that model, and we either empirically measure or estimate the cost of each feature evaluation. Our meta-model deals with the experts' cost by minimizing the overall cost. The algorithm considers besides the classification accuracy of each expert, their cost. 
  
This work is organized as follows: In section \ref{sec:related} we present a brief description of the current research in mixtures of experts and related topics. In section \ref{sec:methodology} we give all the details of the proposed methodology. In section \ref{sec:results} we describe the experimental results obtained in different tests with real datasets. Finally in section \ref{sec:conclusions} we discuss the main results of our work.
\\

\section{Related work}
\label{sec:related}

There are dozens of different methodologies to improve classification rates by combining the ``expertise" of different classifiers. This whole topic is known in the machine learning community as \textit{ensemble learning} or \textit{mixture of experts} \citep{Bishop2006, Jordan1993}. One of the earliest discussions of ensemble learning appears in the work of \citet{Bazell2001}, in which they combine different instances of the same model via bootstrapping, train each classifier in a randomly chosen sub-sample of the training set, and finally predict through majority voting. 

The work from \citet{freund1999short} introduce the \textit{Adaboost} algorithm, which combines classifiers in a cascade scheme. In the cascade of classifiers, each model trains only with the instances that the previous model predicted wrongly. This is achieved by tuning hyperparameters that control the false positive rate and the minimum acceptable detection rate. Unfortunately, this method works only for the two class problem though there are works that discuss extensions to multi-classes \citep{lin2005robust,zehnder2008efficient}.

Although this ensemble method achieves good classification rates, it cannot be applied to our problem. This is because the combination and the classifiers are trained together;  classifiers are not already trained. In our case, experts are already trained, and we do not need to train them again.  On the contrary, we desire to re-use previously acquired knowledge. Moreover, most of boosting methods assume that instances of each model are represented through the same feature space (they use the same features on every model), and the predicted classes for every model are also the same. In our case, we use different features for each model and different output classes.

In \citet{Faraway2014} they have some similarity to our work in the sense that one of the classifiers they consider is hierarchical. They first evaluate if the object is a transient or not, and depending on the answer they attempt to classify among the other classes. What makes a big difference is that we propose a model that automatically learns that hierarchy and able to create different hierarchical classifiers depending on the case.  On the other hand, \citet{Faraway2014} define a hierarchical classifier where the structure is set by hand and  there is no any learning process about that hierarchy.

A seminal work in the mixture of experts was proposed in \citep{jordan1994hierarchical}. They create a hierarchy of base level models that specialize in separate areas of the input space. On each level of the hierarchy, each expert is combined by a gate function that learns a model-combination function that varies depending on the instance to be classified. The combination function assigns a weight to each model in the final prediction. In the original paper, this function is a multinomial distribution, but there exist extensions using probability models from the exponential family \citep{xu1995alternative}. Unfortunately, like most of the current machine learning approaches in mixtures of experts, this method is not helpful for our proposes, because the gate functions and the base models are all trained together to create the effect of specialization/cooperation and therefore all the models must belong to the same problem context (features and classes).

Another perspective of ensemble modelling, in the context of meta-models, is the use of a technique called \textit{Stacked Generalization} \citep{wolpert1992stacked}. In this framework, each base model (or \textit{level-0} model, in the nomenclature of the cited work) is ``fed'' with the data and the output of these models is considered as an input for a meta-model (\textit{level-1} model). This essentially creates an \textit{Intermediate Feature Space} \citep{kuncheva2004combining} where the second stage learning can be performed. In the work of \citet{wolpert1992stacked} this is referred as \textit{level-1} dataset, and the \textit{level-0} dataset is where the \textit{level-0} models are trained. This process can be repeated an indefinite number of times. Intuitively, the meta-model objective is to correct the bias of the base models \citep{leblanc1996combining}. 

Most of the methods mentioned above, rather than focus on work reutilization they concentrate on the ``divide and conquer'' principle and they do not consider cost, making them hard to use in the framework we are addressing in this work. Furthermore, to the best of our knowledge, there are no work in the field of astronomy addressing the problem of automatically learn how to combine previously learned models. We believe that in the area of lightcurve classification, automatic integration can make important contributions, especially with the continuous growth of data and models.


\section{Proposed Method}
\label{sec:methodology}

We start by assuming that we have $m$ already trained models $\{M_1, M_2, \ldots, M_m\}$, where each model $M_i$ correspond to a lightcurve classifier. Each classifier $M_i$
uses a specific set of features $F_{M_i}$ to represent the lightcurves and classifies each lightcurve into a set $C_{M_i}$ of possible classes. For example, $M_0$ can be a model that uses the features $F_{M_0} = \{ Amplitude, Autocor-length, CAR-tau, FluxPercentileRatio\}$, and is able to classify them into  $C_{M_i} = \{ QSO, RRL, Be, Other \}$. Besides having the already trained models $\{M_1, M_2, \ldots, M_m\}$, we have a training set ($D_N$), corresponding to the data associated with the new classification problem, the one we need to solve with the trained models.
\subsection{Creation of the Predictions Dataset}
\label{sec:create_data_pred}

\begin{figure}
	\includegraphics[scale=0.24]{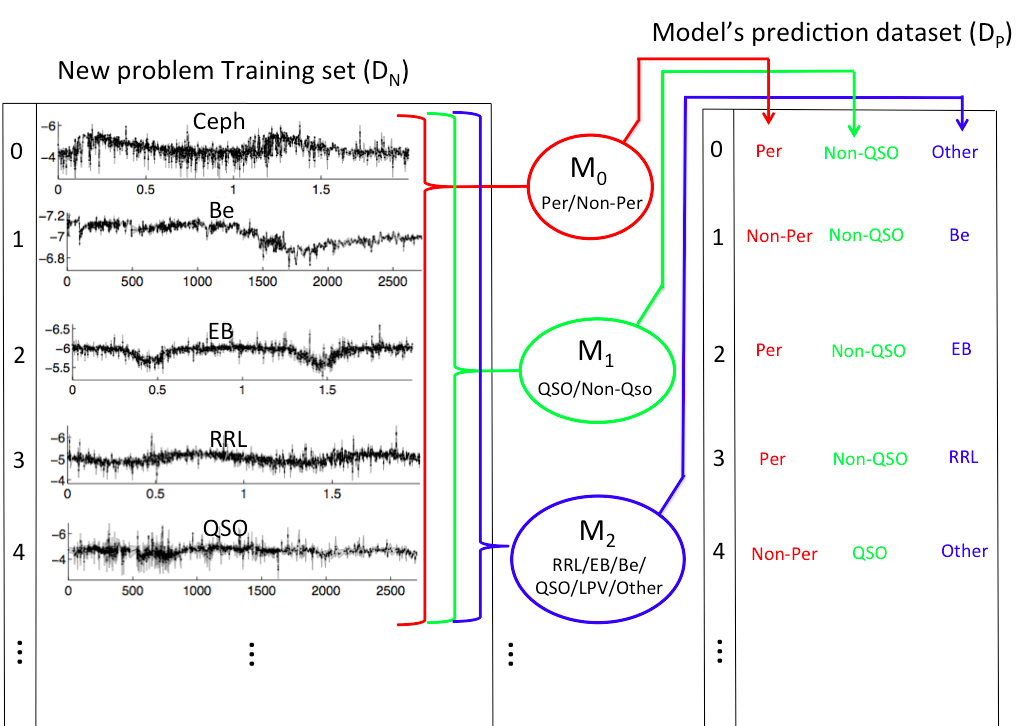}
	\caption{Graphical description of the creation of models predictions data ($D_{P}$). We can see that starting from a set of labeled lightcurves ($D_N$) we ask to each model to predict the class to each of those lightcurves, then we record their predictions as new rows in $D_P$, generating a second dataset later used to train the mixture of models.}
	\label{fig:data_pred}
\end{figure}

The first step of the process is to create a (meta) dataset containing the predictions of each model ($D_{P}$) associated to each of the lightcurves in $D_N$. The main purpose of $D_{P}$ is to have training data for the meta model.
To create $D_{P}$, we just run each of the trained models getting their predictions on the training set $D_N$.  Then, we save those predictions as rows in $D_{P}$ together with the real class label of the lightcurve. Figure \ref{fig:data_pred} shows an example of this process for three given models. 

\subsection{Meta model representation}
\label{sec:meta_repr}

\begin{figure}
	\includegraphics[scale=0.4]{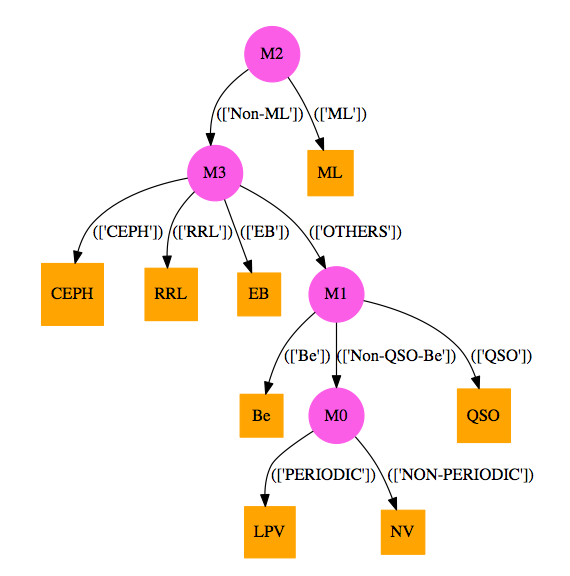}
	\caption{Example of a meta model. Round nodes represent the previously trained models, edges show which path to follow depending on the model´s prediction, square nodes represent the leaves that correspond to a final prediction done by the meta-model.}
	\label{fig:ex_meta}
\end{figure}

 After obtaining $D_P$, we can build a meta-model that efficiently mixes the decision of each of the previously trained models. The meta-model has to act as a ``director''. Every time the meta-model receives a new query lightcurve, it has to choose who is the first model to be used, then depending on the prediction of that model, select the next model, and so forth. A natural representation of the meta-model is a decision tree structured schema, where each node represents one of the previously trained models $M_i$. Each of the edges pointing out from each node represents one of the possible predictions made from the model represented by the node, and leaves represent a final prediction done by the meta-model. Figure \ref{fig:ex_meta} shows an example with four models $\{M_0, M_1, M_2, M_3\}$. The tree structure meta model first asks model $M_2$ to do the prediction. In the case that $M_2$ predicts a microlensing ($ML$), the meta-model immediately predicts $ML$ (reaches a leaf).  In the case that model $M_2$ says $Non-ML$ the meta model asks  model $M_3$ for a prediction. If $M_3$ predicts Cepheid ($CEPH$), RR Lyrae ($RRL$) or Eclipsing Binary ($EB$) the meta-model predicts according to $M_3$, but in the  case that $M_3$ predicts $OTHERS$, the meta-model now asks  $M_1$ for a prediction and so on. This kind of structure is very suitable for what we need. It is very easy to understand, uses a very well-known data structure from computer science (very mature searching and traversal algorithms), and the most important benefit is that it can be interpretable.

\subsection{Automatically building the meta model}
\label{sec:train_meta_model}

  After understanding the structure of the meta-model and how it works, the central question is how we build it? 
We propose an algorithm that is mainly driven by the probability that a given model correctly predicts the class of a lightcurve and the cost of running that model. 
  The likelihood that a model correctly predicts the class of a given lightcurve can be estimated from the training data ($D_P$), and the cost of running that model can be easily calculated from the cost of all the features the model uses to represent the lightcurve.\\ 
  
  The meta-model learning algorithm is inspired by the classical decision tree learning algorithm \citep{Quinlan:1986, Quinlan:1993}. Given a score that measures the quality of any node, the best node is selected to be the root of the tree. Then the algorithm traverse down from each of the possible edges pointing out from the root (possible predictions of the model associated to the root) and recursively searches for the next best model. 
 We select the best model ($M_*$) for a given node of the tree as follows: 
  

\small
\begin{eqnarray}
  M_* = \underset{M_i}{\arg\max} \; \frac{{\rm Info\_Gain}(M_{i})}{E \big[ {\rm Cost}(M_{i}) \big]}  ,  \;  i \in [1, \ldots, m]
\label{eq:best_model}
\end{eqnarray}
\normalsize

\noindent where ${\rm Info\_Gain}(M_{i})$ is the information gain \citep{Quinlan:1986} of model $M_i$, which measures the expected reduction in entropy  in $D_P$ when model $M_i$ makes a prediction. It is defined as: 

\small
\begin{eqnarray}
  {\rm Info\_Gain}(M_i) &=&  H(\rmclass) - \displaystyle \sum_{\substack{v \in \\ C_{M_i}}} \frac{|v|}{|C_{M_i}|} H(\rmclass | v) \nonumber\\
  H(\rmclass) &=& -\sum_{k \in C^M}  \frac{|k|}{|C^M|} \log \frac{|k|}{|C^M|} \nonumber\\
  H(\rmclass | v) &=& -\sum_{k \in C^M_v}  \frac{|k|}{|C^M_v|} \log \frac{|k|}{|C^M_v|}
\end{eqnarray}
\normalsize

\noindent where $C^M$ is the union of all possible classes predicted among all models. Similarly, $C^M_v$ is the union of all classes predicted across the models $\{M_1, M_2, \ldots, M_{i-1}, M_{i+1}, \ldots, M_m\}$ when the model $M_i$ predicts $v$. In simpler words $H(\rmclass | v)$ is the entropy of the class column of $D_P$ selecting only the rows of $D_P$ that match $M_i = v$. Intuitively the information gain tell us if model's $M_i$ predictions are good to separate among possible classes, in the sense that if every time we instantiate the model $M_i$ to it´s possible predictions, we see that the uncertainty in the class column is reduced or not (entropy). This concept is directly related to the probability of getting a successful classification if the meta-model uses $M_i$ to do the final prediction.\\

The term $E \big[ \rmcost(M_{i}) \big]$ is the expected cost of a model, estimated as follows:

\small
\begin{eqnarray}
E \big[ \rmcost(M_{i}) \big] & = & P_{L}(M_{i}) \rmcost(M_{i}) + (1 - P_{L}(M_{i})) \big[ \nonumber \\
&& \displaystyle \sum_{\substack{v \in \\ C_{M_i}}} \sum_{j = i + 1}^{m} P_{L}(M_j | M_i = v)  \times \nonumber\\ 
&& \times \rmcost(M_{j} | M_i = v )\big]\label{eq:E_cost}
\end{eqnarray}

\normalsize

\noindent The term $P_L(M_{i})$ indicates the probability that model $M_i$ reaches a leaf in the tree in the next step.  In other words, how likely is that model $M_i$ will be making a final decision (reaching a leaf). Given that the decision tree algorithm creates a leaf every time most of the remaining instances belong to the same class, to estimate the probability of reaching a leaf we need an indicator of how good was the model after predicting a given class. This is also related to the information gain of the model at that level of the tree.  To have  valid probability values, we normalize the information gain from $[0, 1]$ as:
\small
\begin{eqnarray}
P_{L}(M_{i}) &\approx & 1 - \frac{\displaystyle \sum_{v \in C_{M_i}} \frac{|v|}{|C_{M_i}|} H(\rmclass|v)}{H(\rmclass)}
\end{eqnarray}

The cost of model $M_i$ ($\rmcost(M_{i})$) is calculated as the sum of the features that model $M_i$ uses to represent each lightcurve. The second part of equation \ref{eq:E_cost} is basically the weighted sum of every model but model $M_i$ cost, where each weight corresponds to the probability that the given model reaches a correct leaf in the tree in the next step. Intuitively, equation \ref{eq:best_model} is finding the model whose cost is minimum and at the same time take the meta-model models to the right prediction.\\ 

We summarize the training and predicting steps of the Meta-Classification process below :\\
 
\underline{Training:}
 \begin{itemize}
   \item For each new model $M_i$ create a new data column $D_{P}[i]$ with the prediction of each model $M_i$ over the training data.
   \item Build  $D_{P}$ as a union of all the predictions, $D_{P} = \bigcup_{i=1}^{m} D_{P}[i]$
   \item Create the meta-training set adding to $D_{P}$ a column with the known class of each object (this is just the same class column included in $D_N$).
   \item Build the meta-model according to section \ref{sec:train_meta_model}
  \end{itemize}  
  
  \underline{Predicting:}
   \begin{itemize} 
   \item For any unclassified lightcurve $x$, start traversing the meta-model tree from the root
   \item On each node $M_i$, extract the features $F_{M_i}$, go down the tree according to the prediction of $M_i$ until a leaf is reached.
   \item Predict according to the reached leaf.
 \end{itemize}
 
\section{Experimental Results}
\label{sec:results}
 We tested our model with two lightcurve datasets, MACHO \citep{Cook1995} and EROS-2 \citep{Tisserand:2007}. On each dataset, we created different experts models trained to classify among different subsets variability classes. Each model in the setup uses a specific set of features to describe the lightcurves. These specific sets are determined using a feature importance algorithm called \textit{mean decrease impurity}, described in \citet{Breiman1984}. After a particular model $M_i$ is trained, if the meta model requires a prediction from $M_i$, it only will extract the features included on $M_i$'s specific set. Note that if previously other model extracted some of the required features, they will not be extracted again.
  
 For each of the experts we use a Random Forest classifier \citep{Breiman:2001}. We use  FATS (Feature Analysis for Time Series; \citep{Nun2015}) tool to extract the features of lightcurves. This tool is able to extract up to 64 different features per lightcurve. All details about the meaning of each of the features can be found in \citet{Nun2015}. As mentioned above,  some of the features are more expensive than others. Since each expert uses a selection of the best features according to its own classification problem, models have different associated cost.
 
 
 All the accuracy results are presented throughout recall, precision, F-score and confusion matrix. All these indicators were obtained using a 10-fold cross validation process on each of the training sets. 
 
 \subsection{MACHO dataset}
 
 The MACHO Project (Massive Compact Halo Objects) \citep{Cook1995} observed the Magellanic Clouds and Galactic bulge with the main purpose of detecting microlensing events. Observations were done using blue ($\sim$ 4500 to 6300\AA) and red ($\sim$ 6300 to 7600 \AA) passbands. The cadence is about 1 observation per 2-days for 7.4 years, which generates approximately 1000 observations per object. The lightcurves used in this work are  from the Small and Large Magellanic Clouds. The  fields cover almost the entire LMC bar (10 square degrees) to a limiting magnitude of V $\approx$ 22. The training set contains 6059 labeled lightcurves \citep{Kim:2011}. Table \ref{tab:MACHO_inst_per_class} shows the number of lightcurve per each of the available classes. We created seven models to work as experts, each one trained on a specific problem, with a specific set of features. Table \ref{tab:MACHO_models} shows the features used on each model 
and the classes each model predicts. 
 
\begin{table}[!htbp]
\begin{center}
\begin{tabular}{|cc|}
\hline
class & \# instances\\
\hline
 Be stars & 127\\
CEPH & 101\\
EB & 255\\
LPV & 361\\
ML & 580\\
NV & 3963\\
QSO & 59\\
RRL & 613\\
\hline
\end{tabular}
\caption{Number of instances per each class of variability in MACHO training set}
\label{tab:MACHO_inst_per_class}
\end{center}
\end{table}

\begin{table*}[!hb]
\begin{tabular}{|c|c|c|c|}
\hline
\textbf{Name} & \textbf{Features used in the model} & \textbf{Possible classes} & \textbf{\parbox[t]{1.5cm}{Avg. cost per 
																				lightcurve (secs)}}\\
\hline
$M_0$ & Psi eta, StetsonL, Psi CS, PeriodLS, StetsonJ, Rcs, Period fit, StetsonK AC & PERIODIC, NON-PERIODIC & 1.729\\
\hline
$M_1$ & \parbox[t]{10cm}{Rcs, Color, PeriodLS, Psi CS, Auto-cor-length, Mean, MedianAbsDev, StetsonJ, CAR tau, 
					CAR mean, StetsonL, PercentDifferenceFluxPercentile, Q31, SlottedA length, Eta e, AndersonDarling, 
					Con, FluxPercentileRatioMid65, Freq1 harmonics rel phase 1, Q31 color, Freq2 harmonics amplitude 2, 
					Meanvariance, MedianBRP, Skew, MaxSlope} & NON-QSO, QSO & 2.554\\
\hline
$M_2$ & \parbox[t]{10cm}{Rcs, PeriodLS, Color, Autocor length, Psi CS, SlottedA length, StetsonL, Meanvariance, StetsonJ, 
					PercentAmplitude, Amplitude, Std, Mean, Psi eta, CAR tau, FluxPercentileRatioMid65, Con, 
					Freq3 harmonics amplitude 0} & Non-QSO-Be, Be, QSO & 2.551\\			                    
\hline
$M_3$ & \parbox[t]{10cm}{Color, Con, SlottedA length, Mean, Rcs, StetsonK, Eta e, Skew} & Non-ML, ML & 0.823\\
\hline
$M_4$ & \parbox[t]{10cm}{Psi eta, PeriodLS, Rcs, Psi CS, CAR mean, StetsonL, CAR tau, Period fit, StetsonJ, 
					   FluxPercentileRatioMid35, Skew, Mean, Color} & CEPH, RRL, EB, OTHERS & 1.730\\
\hline
$M_5$ & \parbox[t]{10cm}{Psi eta, SlottedA length, Psi CS, StetsonJ, Color, StetsonL, Period fit, StetsonK AC, Con, Rcs, 
					FluxPercentileRatioMid35, FluxPercentileRatioMid50, Eta e, Skew, Beyond1Std, FluxPercentileRatioMid80, 
					FluxPercentileRatioMid65, FluxPercentileRatioMid20, PeriodLS, MedianBRP} & CEPH, OTHERS, NV, EB & 2.550\\
\hline
$M_6$ & \parbox[t]{10cm}{Color, Rcs, Skew, SlottedA length, Con, Psi CS, Psi eta, StetsonJ, PeriodLS, Eta e, StetsonK, FluxPercentileRatioMid35, 
					Mean, Period fit, CAR mean, StetsonL, FluxPercentileRatioMid50, FluxPercentileRatioMid20, FluxPercentileRatioMid65, 
					CAR tau, Autocor length, Q31 color, Beyond1Std} & EB, OTHERS, Be, ML & 2.552\\
\hline
\end{tabular}\\
\caption{Pre-trained models for MACHO dataset, features used on each model, classes that each model can predict and cost that each model takes to represent one lightcurve. The cost is directly related with the features models need to
extract in order to classify a given lightcurve.}
\label{tab:MACHO_models}
\end{table*}

Table \ref{tab:acc_experts_MACHO} presents the precision, recall and f-score of each of the classes per each of the models. Most of the models are getting high f-score for all their classes. We can see that quasars are the most complicated objects, mainly because they are confused with Be stars (model $M2$). 

\begin{table}[H]
\begin{center}
\begin{tabular}{|ccccc|}
\hline
&&&&\\
\textbf{Model} & \textbf{Class} & \textbf{Precision} & \textbf{Recall} & \textbf{F-Score}\\
&&&&\\
\hline
     M6      &      Be      &   0.733     &    0.780     &    0.756    \\
             &    OTHERS    &   0.985     &    0.985     &    0.985    \\
             &      ML      &   0.970     &    0.964     &    0.967    \\
             &      EB      &   0.877     &    0.867     &    0.872    \\
\hline
     M0      &   PERIODIC   &   0.957     &    0.962     &    0.960    \\
             & NON-PERIODIC &   0.989     &    0.988     &    0.989    \\
\hline
     M3      &    Non-ML    &   0.995     &    0.997     &    0.996    \\
             &      ML      &   0.970     &    0.957     &    0.964    \\
\hline
     M2      &      Be      &   0.866     &    0.811     &    0.837    \\
             &     QSO      &   0.738     &    0.525     &    0.614    \\
             &  Non-QSO-Be  &   0.994     &    0.998     &    0.996    \\
\hline
     M4      &     CEPH     &   0.929     &    0.901     &    0.915    \\
             &     RRL      &   0.967     &    0.949     &    0.958    \\
             &      EB      &   0.900     &    0.878     &    0.889    \\
             &    OTHERS    &   0.993     &    0.997     &    0.995    \\
\hline
     M1      &   NON-QSO    &   0.995     &    0.998     &    0.996    \\
             &     QSO      &   0.675     &    0.458     &    0.545    \\
\hline
     M5      &     CEPH     &   0.936     &    0.871     &    0.903    \\
             &    OTHERS    &   0.954     &    0.976     &    0.965    \\
             &      EB      &   0.897     &    0.855     &    0.876    \\
             &      NV      &   0.992     &    0.987     &    0.990    \\
\hline
\end{tabular}\\
\caption{Accuracy indicators per each class on each of the models problem, in MACHO dataset.}
\label{tab:acc_experts_MACHO}
\end{center}
\end{table}

\begin{figure}
\centering
	\includegraphics[scale=0.4]{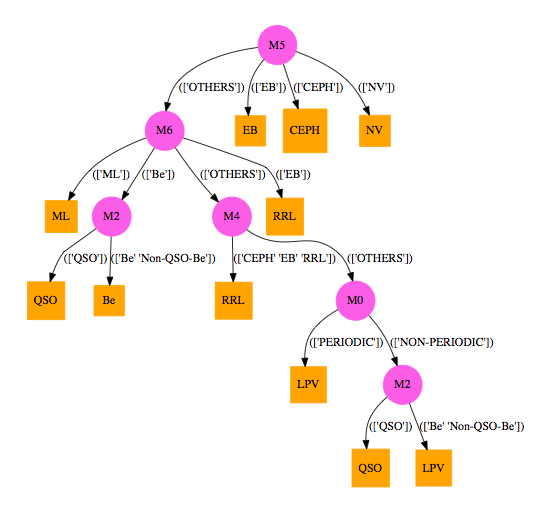}
	\caption{Meta-model learned from the MACHO training set.}
	\label{fig:meta_model_MACHO}
\end{figure}

After learning the meta model from the MACHO data using the proposed algorithm, we obtained the structure that is shown in Figure \ref{fig:meta_model_MACHO}. We can see how the meta-model performs the classification. The meta-model starts by asking $M5$ and if $M5$ predicts ``EB'', ``CEPH'' or ``NV'', the meta-model predicts as $M5$ without asking any other model, but if $M5$ predicts ``Others'', then the meta-model asks for the prediction from $M6$, and so on. From the tree, we can also see in most cases that the meta-model asks other models when the prediction is not so confident, like ``Others'' or when there is a hard class. For example, if $M6$ says that the object is a Be star, the meta-model does not predict immediately, but it continues and asks  $M2$, which also knows about Be stars, and  $M2$ predicts a Quasar (which usually is confused with Be stars); the meta-model also predicts a Quasar. If $M2$  predicts a Be star, the meta-model can also predict a Be star. A more interesting situation occurs when $M2$ predicts ``Non-QSO-Be'' (Non-Quasar and Non-Be star). In this case, the best decision the meta-model can do is to predict a Be star, which is more likely than any other class given that $M6$ predicted a Be star. It is also interesting to see that the meta-model can classify Long Period Variables (LPV) even if none of the previous models can classify them, mainly because from the training set the meta-model could infer prediction patterns from the models that occur together with the LPV class. From the tree, we can see that the meta-model is predicting LPV by discarding the other classes because most of the edges along the paths that end up in an LPV tree correspond to predictions for ``Others''  or ``Non-$<$some classes$>$" from most of the models. Another very fascinating pattern happens with LPV; the meta-model realizes that LPV is a Periodic star, after the model $M0$ in the 4th level of the tree. Note also the meta-model does not use  model $M1$. Model $M1$ classifies between Non-QSO and QSO, which are classes already covered by model $M2$.

\begin{figure}[H]
\centering
	\includegraphics[scale=0.35]{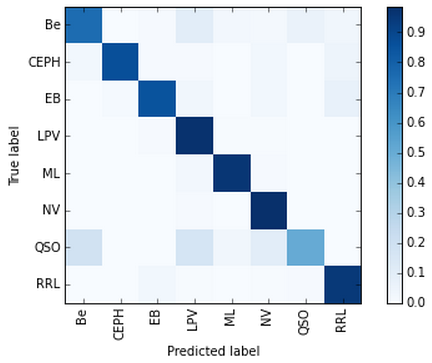}
	\caption{Confusion matrix for the meta-model learned from the MACHO training set.}
	\label{fig:cm_meta_MACHO}
\end{figure}

 To show that the meta model does not sacrifice performance after the integration, Table \ref{tab:acc_meta_MACHO} shows recall, precision and F-score of the final meta model. 
 and in Figure \ref{fig:cm_meta_MACHO} we can see the confusion matrix of the meta-classifier. Most of recall, precision and f-score values are maintained in the meta-classifier, even some indicators are improving, like in the case of Cepheids, as a result from the collaboration of two different models that are able to classify Cepheids. 

\begin{table}[H] 
\begin{center}
\begin{tabular}{|cccc|}
\hline
&&&\\
\textbf{Class} & \textbf{Precision} & \textbf{Recall} & \textbf{F-Score}\\
&&&\\
\hline
     Be      &    0.857     &    0.756     &    0.803    \\
\hline
    CEPH     &    0.936     &    0.871     &    0.903    \\
\hline
     EB      &    0.897     &    0.855     &    0.876    \\
\hline
    LPV      &    0.799     &    0.978     &    0.879    \\
\hline
     ML      &    0.977     &    0.960     &    0.969    \\
\hline
     NV      &    0.992     &    0.987     &    0.990    \\
\hline
    QSO      &    0.732     &    0.508     &    0.600    \\
\hline
    RRL      &    0.946     &    0.949     &    0.948    \\
\hline
\end{tabular}\\
\caption{Accuracy indicators per each class for the meta-model, in MACHO training set.}
\label{tab:acc_meta_MACHO}
\end{center}
\end{table}

\begin{figure}
\centering
	\includegraphics[scale=0.4]{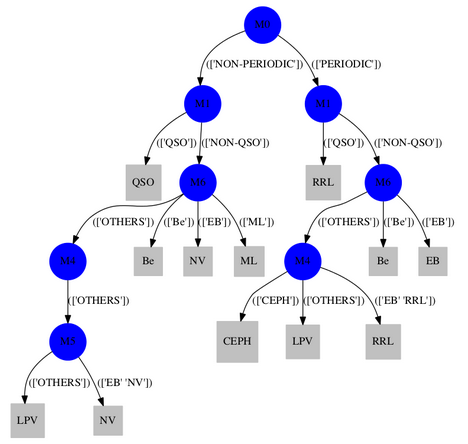}
	\caption{Meta-model learned from the MACHO training set without considering the of the models.}
	\label{fig:meta_model_MACHO_No_Cost}
\end{figure}

\begin{figure}[H]
\centering
	\includegraphics[scale=0.35]{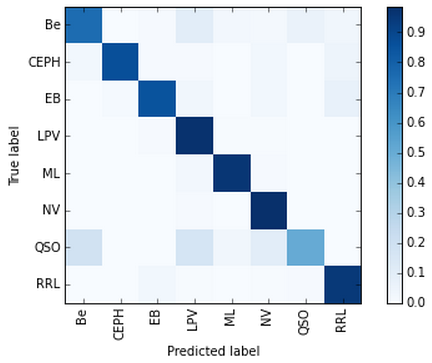}
	\caption{Confusion matrix for the meta-model without considering the cost of the experts, learned from the MACHO training set.}
	\label{fig:cm_meta_MACHO_no_cost}
\end{figure}

  To show the contribution of the cost estimation of each model, we run the same experiment just considering the information gain in the score of each model, in other words, we assume that all models have the same cost. The resulting meta-model is shown in Figure \ref{fig:meta_model_MACHO_No_Cost}. The meta-model, in this case, is less efficient, asking for a prediction more than once to most of the models, for example, independently of the prediction of model $M0$, the meta model asks twice for a prediction to $M1$. Also, note that this meta-model decides to use $M1$ instead of $M2$, which is a cheaper model but not necessarily worst than $M1$. As we can see from Table \ref{tab:acc_meta_MACHO_No_Cost} and confusion matrix in Figure \ref{fig:cm_meta_MACHO_no_cost}, there is no strong difference between the classification results; only in Cepheids we can see a 2\% of improvement in the F-score when the meta-classifier does not penalize each model according to their cost, but there is a drop in F-score for the class Be stars. Calculating the total training cost for the meta-classifier in both cases (with and without considering the cost of the expert models), when the meta-model does not take into account the cost, the training process takes 167\% longer than in the case when the meta-classifier takes into account the cost of the model experts. 

\begin{table}[H] 
\begin{center}
\begin{tabular}{|cccc|}
\hline
&&&\\
\textbf{Class} & \textbf{Precision} & \textbf{Recall} & \textbf{F-Score}\\
&&&\\
\hline
     Be      &    0.832     &    0.740     &    0.783    \\
\hline
    CEPH     &    0.938     &    0.901     &    0.919    \\
\hline
     EB      &    0.884     &    0.863     &    0.873    \\
\hline
    LPV      &    0.779     &    0.978     &    0.867    \\
\hline
     ML      &    0.974     &    0.964     &    0.969    \\
\hline
     NV      &    0.993     &    0.985     &    0.989    \\
\hline
    QSO      &    0.667     &    0.441     &    0.531    \\
\hline
    RRL      &    0.954     &    0.938     &    0.946    \\
\hline
\end{tabular}\\
\caption{Accuracy indicators per each class for the meta-model without considering the cost of models, in MACHO training set.}
\label{tab:acc_meta_MACHO_No_Cost}
\end{center}
\end{table}

 \subsection{EROS dataset}
The EROS project (Exp\'erience de Recherche d’Objets Sombres) \citep{Derue1999} observed the Galactic Spiral Arms (GSA), LMC, SMC and Galactic bulge during 6.7 years,
dedicated to detect microlensing events. Observations were done in two non standard passbands. One is the EROS-red passband $R_{\rm E}$, centred on  $\bar\lambda = 762\ {\rm nm}$ and EROS-visible passband $V_{\rm E}$, centred on  $\bar\lambda = 600\ {\rm nm}$. The lightcurves used in this work are from the LMC (60 fields) and SMC (10 fields). The limiting magnitud of the EROS $V_{\rm E}$ band is $\sim$ 20. The cadence varies among the fields, but in average about 500 observations were obtained for each lightcurve. The training set contains 68,718 labeled lightcurves, obtained from \citet{Kim2014}. Table \ref{tab:EROS_inst_per_class} shows the number of lightcurves per each of the available classes. This training set is more complex than MACHO training set, in the sense that some subclasses of variability are added to the problem, making the separation more challenging due to the similarity among some classes. Our main goal is not to solve the classification problem for all the subclasses but to solve the integration problem using the provided expert models.  Therefore, in cases where the respective experts do not well classify some subclasses, the meta-model will probably not be able to classify well those classes either. We used six model experts, each one trained on a specific problem, with a specific set of features. Table \ref{tab:EROS_models} shows the features used on each model, the available classes each model can predict and the average cost per lightcurve that the model takes to perform classification.  
\begin{table}[!htbp]
\begin{center}
\begin{tabular}{|cc|}
\hline
class & \# instances\\
\hline
Ceph 1O & 870\\
Ceph F & 1272\\
Ceph 1O 2O & 111\\
EB & 13523\\
LPV OSARG RGB O & 31487\\
LPV SRV AGB O & 4337\\
LPV SRV AGB C & 3748\\
LPV Mira AGB C & 760\\
LPV Mira AGB O & 320\\
RRL & 12167\\
T2CEPH & 123\\
\hline
\end{tabular}
\caption{Number of instances per each class of variability in EROS training set.}
\label{tab:EROS_inst_per_class}
\end{center}
\end{table}

\begin{table*}[!hb]
\begin{tabular}{|c|c|c|c|}
\hline
\textbf{Name} & \textbf{Features used in the model} & \textbf{Possible classes} & \textbf{\parbox[t]{1.5cm}{Avg. cost per 
																				lightcurve (secs)}}\\
\hline
$M_0$ & \parbox[t]{9cm}{Color, Mean, PeriodLS, CAR mean, Q31 color, Autocor length, CAR tau, FluxPercentileRatioMid50, FluxPercentileRatioMid35, 
					SlottedA length, FluxPercentileRatioMid65, Rcs, Q31, MedianAbsDev, FluxPercentileRatioMid20, Beyond1Std, CAR sigma, 
					Amplitude, Psi eta} & \parbox[t]{4cm}{OTHERS, CEPHEID, RRL, EB} & 16.326\\
\hline
$M_1$ & \parbox[t]{9cm}{Color, Mean, CAR mean, Autocor length, Q31 color, CAR tau, SlottedA length, PeriodLS, Rcs, CAR sigma, Amplitude, 
					PercentDifferenceFluxPercentile} & \parbox[t]{4cm}{LPV, Non-LPV} & 16.325\\
\hline
$M_2$ & \parbox[t]{9cm}{Color, Mean, PeriodLS, CAR mean, Q31 color, CAR tau, FluxPercentileRatioMid50, SlottedA length, FluxPercentileRatioMid35, 
					Autocor length, FluxPercentileRatioMid65, FluxPercentileRatioMid20, CAR sigma, Rcs, Q31, MedianAbsDev, Beyond1Std, 
					Amplitude, Freq1 harmonics amplitude 0, Psi eta} & \parbox[t]{4cm}{OTHERS, RRL, CEPH T2CEPH, EB} & 21.644\\			                    
\hline
$M_3$ & \parbox[t]{9cm}{PeriodLS, Mean, Color, FluxPercentileRatioMid50, Q31, FluxPercentileRatioMid35, Q31 color, MedianAbsDev, CAR mean, 
					FluxPercentileRatioMid65, CAR tau, Freq1 harmonics amplitude 0, FluxPercentileRatioMid20, Beyond1Std, Std, StetsonK, 
					Rcs, SlottedA length, StetsonL, FluxPercentileRatioMid80, 
					PercentDifferenceFluxPercentile, Amplitude} & \parbox[t]{4cm}{OTHERS, Ceph 1O 2O, RRL, Ceph 1O, T2CEPH, Ceph F} & 7.866\\
\hline
$M_4$ & \parbox[t]{9cm}{Color, Mean, Q31 color, SlottedA length, PercentDifferenceFluxPercentile, CAR mean, Amplitude, Autocor length, Q31, Std, 
					CAR tau, MedianAbsDev, Meanvariance, StetsonJ, Freq1 harmonics amplitude 0, 
					PeriodLS, Rcs} & \parbox[t]{4cm}{OTHERS, LPV OSARG RGB O, LPV Mira AGB C, LPV SRV AGB O, LPV SRV AGB C, LPV Mira AGB O} & 7.866\\
\hline
$M_5$ & \parbox[t]{9cm}{PeriodLS, Mean, Color, FluxPercentileRatioMid50, FluxPercentileRatioMid35, MedianAbsDev, Q31 color, Q31, CAR mean, Freq1 harmonics amplitude 0, 
					FluxPercentileRatioMid20, FluxPercentileRatioMid65, CAR tau, Beyond1Std, Autocor length, 
					FluxPercentileRatioMid80, SlottedA length, StetsonK, Std, Skew, StetsonL} & \parbox[t]{4cm}{OTHERS, CEPHEID, RRL} & 7.867\\
\hline
\end{tabular}\\
\caption{Pre-trained models for EROS dataset, features used on each model, classes that each model can predict and cost that each model takes to represent one lightcurve. The cost is directly related with the features models need to extract in order to classify a given lightcurve.}
\label{tab:EROS_models}
\end{table*}

Table \ref{tab:acc_experts_EROS} shows the precision, recall and f-score of each of the classes per model. As we can see, in some cases the experts  failed to classify some of the classes. For example, $M3$ it not able to successfully classify T2 Cepheids and also the f-score for Cepheids 1O 2O is lower than the average score of the other models and classes.
This setup, in particular, is showing us that some of the variability classes can not be automatically classified by the expert, making the meta-model learning process harder than the setup with MACHO dataset.

\begin{table}
\begin{center}
\begin{tabular}{|ccccc|}
\hline
&&&&\\
\textbf{Model} & \textbf{Class} & \textbf{Precision} & \textbf{Recall} & \textbf{F-Score}\\
&&&&\\
\hline
     M5      &     RRL      &   0.960     &    0.938     &    0.949    \\
             &    OTHERS    &   0.985     &    0.991     &    0.988    \\
             &   CEPHEID    &   0.962     &    0.952     &    0.957    \\
\hline
     M1      &     LPV      &   0.992     &    0.996     &    0.994    \\
             &   Non-LPV    &   0.994     &    0.988     &    0.991    \\
\hline
     M0      &     RRL      &   0.965     &    0.937     &    0.951    \\
             &   CEPHEID    &   0.957     &    0.957     &    0.957    \\
             &    OTHERS    &   0.992     &    0.995     &    0.993    \\
             &      EB      &   0.939     &    0.954     &    0.946    \\
\hline
     M3      &     RRL      &   0.957     &    0.938     &    0.947    \\
             &    T2CEPH    &   0.944     &    0.545     &    0.691    \\
             &  Ceph 1O 2O  &   0.758     &    0.676     &    0.714    \\
             &   Ceph 1O    &   0.926     &    0.860     &    0.892    \\
             &    Ceph F    &   0.965     &    0.965     &    0.965    \\
             &    OTHERS    &   0.984     &    0.991     &    0.988    \\
\hline
     M2      &     RRL      &   0.966     &    0.938     &    0.952    \\
             & CEPH T2CEPH  &   0.988     &    0.948     &    0.968    \\
             &    OTHERS    &   0.992     &    0.997     &    0.994    \\
             &      EB      &   0.940     &    0.956     &    0.948    \\
\hline
     M4      & LPV Mira AGB C &   0.872     &    0.867     &    0.869    \\
             & LPV SRV AGB C &   0.947     &    0.921     &    0.934    \\
             & LPV OSARG RGB O &   0.974     &    0.989     &    0.982    \\
             & LPV SRV AGB O &   0.901     &    0.863     &    0.882    \\
             &    OTHERS    &   0.994     &    0.989     &    0.992    \\
             & LPV Mira AGB O &   0.904     &    0.794     &    0.845    \\
\hline
\end{tabular}
\caption{Accuracy indicators per each class on each of the models problem, in EROS dataset.}
\label{tab:acc_experts_EROS}
\end{center}
\end{table}

\begin{figure*}
\centering
	\includegraphics[scale = 0.4]{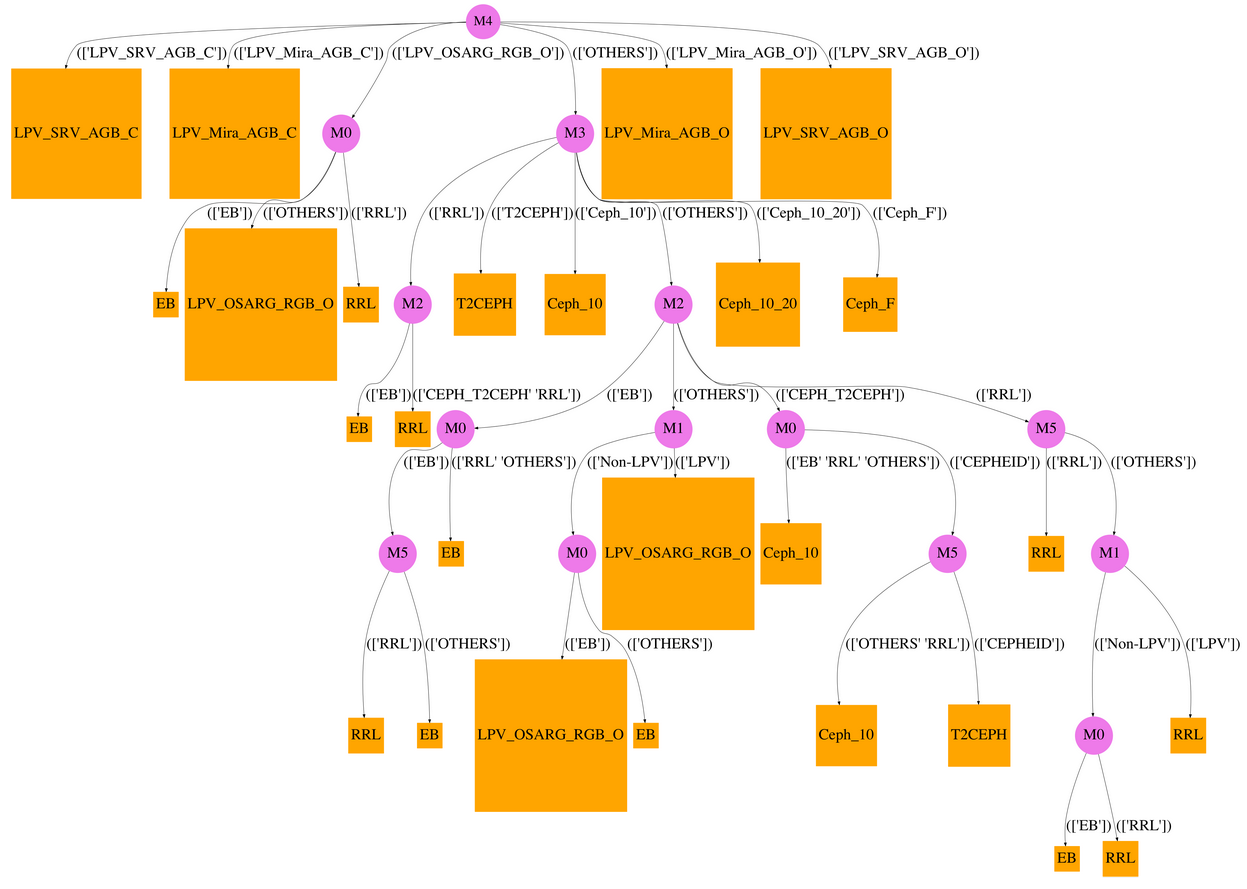}
	\caption{Big picture of the Meta-model learned from the EROS training set.}
	\label{fig:meta_model_EROS}
\end{figure*}

Figure \ref{fig:meta_model_EROS} shows the resulting meta-model for EROS training set. We can see that at the root level, the meta-model asks $M4$ for a classification, in cases where $M4$ predicts LPV-SRV-AGB-C, LPV-Mira-AGB-C, LPV-Mira-AGB-O and LPV-SRV-AGB-O, the meta-model believes $M4$, in other cases it asks for other predictions. That makes sense because $M4$ is the only model trained to separate the subclasses of LPVs. In some cases the
meta-model wants to get more confident about the prediction of some of the LPV subclasses, asking other models and predicting again the LPV subclasses when most of the other models predict ``Others''. When $M4$ predicts LPV-OSARG-RGB-O, the meta-model asks more before taking a final decision. For example asking to $M0$, and in cases where $M0$ is not so confident about one of its classes and predicts ``Others'', the meta-model predicts according to $M4$. Other model that contributes with extra information about the LPV stars is $M1$, which can predict between ``LPV'' or ``Not-LPV''. We can see from the tree that in some cases the meta-model ends up predicting a subclass of LPV´s after most of the models predict ``Others'' and $M1$ predicts a LPV. It is interesting to see how the meta-model takes advantage of having more experts trained to classify RR Lyrae stars. For example, after $M4$ predicts ``Others'', the meta-models asks for $M3$, and when $M3$ predicts a RR Lyrae, instead of immediately believing to it, the meta-model asks to $M2$, and again if $M2$ predicts a RR Lyrae, the meta model also predicts RR Lyrae. More interesting is when $M4$ says ``Others'' and $M3$ also says ``Others'', if $M2$ predicts a RR Lyrae, the meta-model instead of immediately believe to $M2$ asks for a prediction to $M5$, and if $M5$ confirms that it is a RR Lyrae, then the meta-model also predicts a RR Lyrae.

\begin{figure}[H]
\centering
	\includegraphics[scale=0.4]{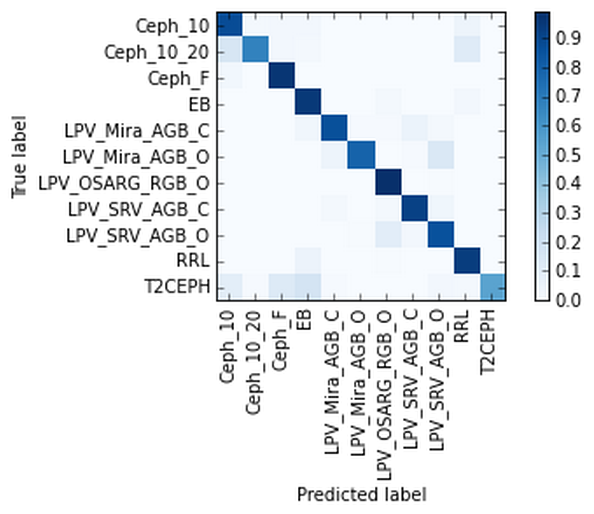}
	\caption{Confusion matrix for the meta-model learned from the EROS training set.}
	\label{fig:cm_meta_EROS}
\end{figure}

 To show that the meta model does not sacrifice performance after the integration, Table \ref{tab:acc_meta_EROS} shows recall, precision and F-score of the final meta model. 
 In Figure \ref{fig:cm_meta_EROS} we can see the confusion matrix of the meta-classifier. Most of recall, precision and f-score values are maintained in the meta-classifier. 

\begin{table}
\begin{center}
\begin{tabular}{|cccc|}
\hline
&&&\\
\textbf{Class} & \textbf{Precision} & \textbf{Recall} & \textbf{F-Score}\\
&&&\\
\hline
  Ceph 1O    &    0.901     &    0.878     &    0.889    \\
\hline
 Ceph 1O 2O  &    0.758     &    0.676     &    0.714    \\
\hline
   Ceph F    &    0.965     &    0.965     &    0.965    \\
\hline
     EB      &    0.938     &    0.954     &    0.946    \\
\hline
LPV Mira AGB C &    0.872     &    0.867     &    0.869    \\
\hline
LPV Mira AGB O &    0.904     &    0.794     &    0.845    \\
\hline
LPV OSARG RGB O &    0.974     &    0.990     &    0.982    \\
\hline
LPV SRV AGB C &    0.947     &    0.921     &    0.934    \\
\hline
LPV SRV AGB O &    0.901     &    0.863     &    0.882    \\
\hline
    RRL      &    0.965     &    0.938     &    0.951    \\
\hline
   T2CEPH    &    0.893     &    0.545     &    0.677    \\
\hline
\end{tabular}\\
\caption{Accuracy indicators per each class for the meta-model, in EROS training set.}
\label{tab:acc_meta_EROS}
\end{center}
\end{table}

\begin{figure*}
\centering
	\includegraphics[scale=0.4]{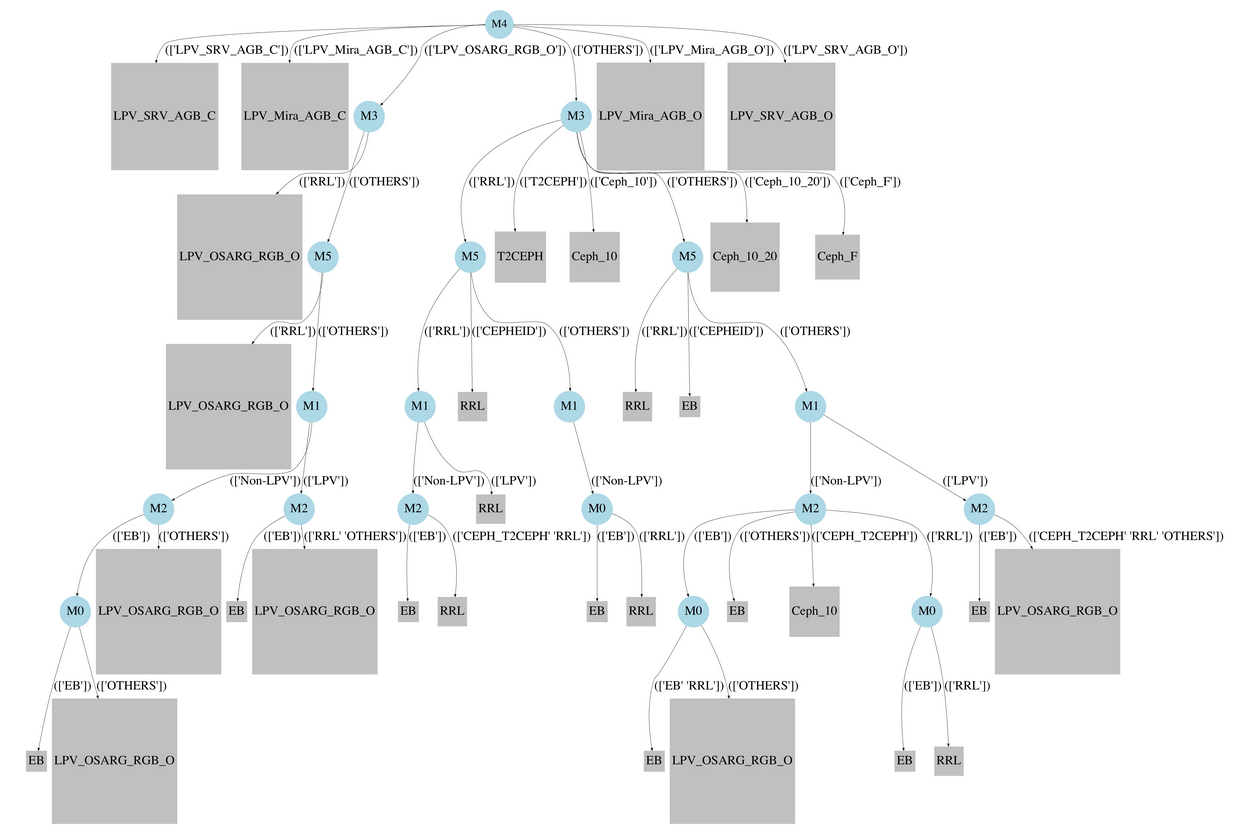}
	\caption{Meta-model learned from the EROS training set without considering the of the models.}
	\label{fig:meta_model_EROS_No_Cost}
\end{figure*}

\begin{figure}
\centering
	\includegraphics[scale=0.4]{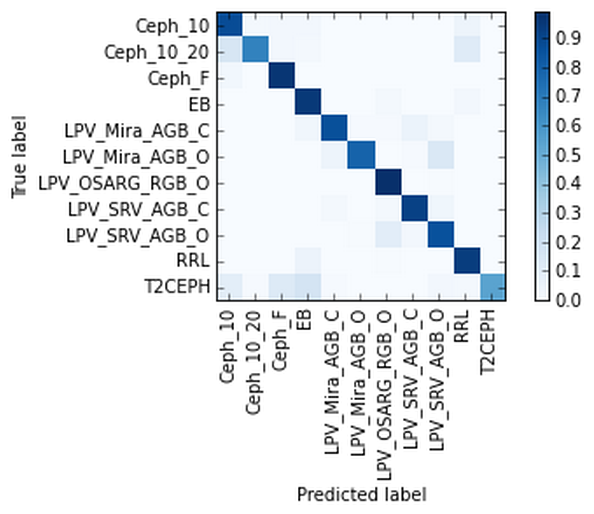}
	\caption{Confusion matrix for the meta-model without considering the cost of the experts, learned from the EROS training set.}
	\label{fig:cm_meta_EROS_no_cost}
\end{figure}

  As we did in the MACHO experiment, in EROS we also run the same experiment without considering the cost of each model. The resulting meta-model is shown in Figure \ref{fig:meta_model_EROS_No_Cost}. Similarly to that in the MACHO case, the meta-model asks many times for a prediction to most of the models, trying to maximize the confidence about the prediction instead of counting how expensive is the process. 
  The meta-model basically asks all the models that can contribute with some information about certain classification, maximizing the confidence without restriction in the number of questions it does. We can see for example that models $M3$ and $M5$ are the most expensive models (Table \ref{tab:EROS_models}), so the meta-model that takes into account the cost, does not call to $M3$ and $M5$ as much as the
   meta-model without cost does. From Table \ref{tab:acc_meta_EROS_No_Cost} and confusion matrix in Figure \ref{fig:cm_meta_EROS_no_cost}, we can see that there is no significative improvement in f-score in the meta-model that does not consider the cost. Calculating the total training cost for the meta-classifier in both cases (with and without considering the cost of the expert models), when the meta-model does not take into account the cost, the training process takes about 80\% longer than in the case when the meta-classifier takes into account the cost of the model experts.

\begin{table}
\begin{center}
\begin{tabular}{|cccc|}
\hline
&&&\\
\textbf{Class} & \textbf{Precision} & \textbf{Recall} & \textbf{F-Score}\\
&&&\\
\hline
  Ceph 1O    &    0.910     &    0.876     &    0.893    \\
\hline
 Ceph 1O 2O  &    0.758     &    0.676     &    0.714    \\
\hline
   Ceph F    &    0.962     &    0.965     &    0.963    \\
\hline
     EB      &    0.936     &    0.955     &    0.945    \\
\hline
LPV Mira AGB C &    0.872     &    0.866     &    0.869    \\
\hline
LPV Mira AGB O &    0.904     &    0.794     &    0.845    \\
\hline
LPV OSARG RGB O &    0.973     &    0.990     &    0.982    \\
\hline
LPV SRV AGB C &    0.947     &    0.921     &    0.934    \\
\hline
LPV SRV AGB O &    0.901     &    0.863     &    0.882    \\
\hline
    RRL      &    0.966     &    0.936     &    0.951    \\
\hline
   T2CEPH    &    0.931     &    0.545     &    0.687    \\
\hline
\end{tabular}\\
\caption{Accuracy indicators per each class for the meta-model without considering the cost of models, in EROS training set.}
\label{tab:acc_meta_EROS_No_Cost}
\end{center}
\end{table}

\section{Conclusions}
\label{sec:conclusions}

We present a novel algorithm that allows astronomers to solve new classification problems by reusing previously trained classifiers. This kind of solutions facilitate a faster development of automated classification methodologies, avoiding to re-train new models from scratch. Upcoming surveys like LSST \citep{Ivezic2008} will demand this kind of solutions, since the amount of data will not allow scientists to waste time re-calibrating models every time new science problems appear. Our intuition is that when a new variable star classification problem arises, if there are classes of stars and features already involved in previous problems, we should be able to use those models in the building process of the new solution. So far most of the research done in the automatic classification of variable stars field show strong relationships among the classes they study and the features they use, any of those classifiers could be plugged into our meta algorithm and be used to build a new solution. An important contribution of this work relies in the possibility to work with different contexts, something that is very natural when model integration occurs; every model has to deal with its own classes and its own data representation, which makes the integration more challenging. So far we have very promising results, the accuracy of the meta-model was as good as the accuracy of the model experts, that is the first goal that a integration model has to achieve.

Another important contribution is that the meta-model is human readable, we can easily observe the meta-model structure, directly inferring how the meta-model acts on every possible situations, making the meta-model more trustable for scientists. As future research we aim to work on the integration of data coming from different kind of telescopes, this creates new challenges to overcome, such as the identification of hidden patterns that come from instrumental differences, and the application of those patterns to the classification models to make them able to work on heterogeneous data. We strongly believe that making efforts in that direction will have a huge impact in the astronomical community. An issue that is not addressed in this work is the fact that the training sets are unbalanced and not properly evaluated. Analysing and generating better training sets is a future research direction. As a matter of fact, there are no good descriptions on how most of the training sets were generated in the first place. For this work, we assume the training sets are given. Fortunately, from the results we can see that Random Forest classifier can deal with unbalanced training sets. The k-fold cross validation process we use is stratified, ensuring that the testing and training sets are created with the same proportions of stars as the initial variability classes.

\newpage

\bibliography{library,paper-meta}

\begin{thebibliography}{46}
\providecommand{\natexlab}[1]{#1}
\providecommand{\url}[1]{\texttt{#1}}
\expandafter\ifx\csname urlstyle\endcsname\relax
  \providecommand{\doi}[1]{doi: #1}\else
  \providecommand{\doi}{doi: \begingroup \urlstyle{rm}\Url}\fi

\bibitem[Bazell and Aha(2001)]{Bazell2001}
D.~Bazell and David W. Aha.
\newblock {Ensembles of Classifiers for Morphological Galaxy Classification},
  2001.
\newblock ISSN 0004-637X.

\bibitem[Belokurov et~al.(2003)Belokurov, Evans, and Du]{Belokurov2003}
V.~Belokurov, N.~W. Evans, and Y.~L. Du.
\newblock {Light-curve classification in massive variability surveys -- I.
  Microlensing}.
\newblock \emph{Monthly Notices of the Royal Astronomical Society},
  341\penalty0 (4):\penalty0 1373--1384, jun 2003.
\newblock ISSN 0035-8711.
\newblock \doi{10.1046/j.1365-8711.2003.06512.x}.
\newblock URL \url{http://arxiv.org/abs/astro-ph/0211121}.

\bibitem[Bishop(2006)]{Bishop2006}
Christopher~M Bishop.
\newblock \emph{{Pattern Recognition and Machine Learning}}, volume~4.
\newblock 2006.
\newblock ISBN 9780387310732.
\newblock \doi{10.1117/1.2819119}.
\newblock URL \url{http://www.library.wisc.edu/selectedtocs/bg0137.pdf}.

\bibitem[Bishop and Svensen(2012)]{Bishop2012}
Christopher~M. Bishop and Markus Svensen.
\newblock {Bayesian Hierarchical Mixtures of Experts}.
\newblock oct 2012.
\newblock URL \url{http://arxiv.org/abs/1212.2447}.

\bibitem[Bloom and Richards(2011)]{Bloom:2011}
J.{\~{}}S. Bloom and J.{\~{}}W. Richards.
\newblock {Data Mining and Machine-Learning in Time-Domain Discovery
  {\{}$\backslash$amp{\}} Classification}.
\newblock \emph{ArXiv e-prints}, 2011.

\bibitem[Bloom et~al.(2011)Bloom, Richards, Nugent, Quimby, Kasliwal, Starr,
  Poznanski, Ofek, Cenko, Butler, Kulkarni, Gal-Yam, and Law]{Bloom2:2011}
J.{\~{}}S. Bloom, J.{\~{}}W. Richards, P.{\~{}}E. Nugent, R.{\~{}}M. Quimby,
  M.{\~{}}M. Kasliwal, D.{\~{}}L. Starr, D~Poznanski, E.{\~{}}O. Ofek,
  S.{\~{}}B. Cenko, N.{\~{}}R. Butler, S.{\~{}}R. Kulkarni, A~Gal-Yam, and
  N~Law.
\newblock {Automating Discovery and Classification of Transients and Variable
  Stars in the Synoptic Survey Era}.
\newblock \emph{ArXiv e-prints}, 2011.

\bibitem[Breiman(2001)]{Breiman:2001}
L~Breiman.
\newblock {Random Forests}.
\newblock In \emph{Machine Learning}, pages 5--32, 2001.

\bibitem[Breiman et~al.(1984)Breiman, Friedman, Olshen, and Stone]{Breiman1984}
L~Breiman, J~H Friedman, R~A Olshen, and C~J Stone.
\newblock \emph{{Classification and Regression Trees}}, volume~19.
\newblock 1984.
\newblock ISBN 0412048418.

\bibitem[Butler and Bloom(2011)]{Butler2011AJ}
N.{\~{}}R. Butler and J.{\~{}}S. Bloom.
\newblock {Optimal Time-series Selection of Quasars}.
\newblock \emph{AJ}, 141:\penalty0 93----+, mar 2011.
\newblock \doi{10.1088/0004-6256/141/3/93}.

\bibitem[Chamroukhi(2015)]{Chamroukhi2015}
Faicel Chamroukhi.
\newblock {Non-Normal Mixtures of Experts}.
\newblock page~61, jun 2015.
\newblock URL \url{http://arxiv.org/abs/1506.06707}.

\bibitem[Cook et~al.(1995)Cook, Alcock, Allsman, Axelrod, Freeman, Peterson,
  Quinn, Rodgers, Bennett, Reimann, Griest, Marshall, Pratt, Stubbs,
  Sutherland, and Welch]{Cook1995}
K.~H. Cook, C.~Alcock, R.~A. Allsman, T.~S. Axelrod, K.~C. Freeman, B.~A.
  Peterson, P.~J. Quinn, A.~W. Rodgers, D.~P. Bennett, J.~Reimann, K.~Griest,
  S.~L. Marshall, M.~R. Pratt, C.~W. Stubbs, W.~Sutherland, and D.~Welch.
\newblock {Variable Stars in the MACHO Collaboration Database}.
\newblock page~10, may 1995.
\newblock URL \url{http://arxiv.org/abs/astro-ph/9505124}.

\bibitem[Derue et~al.(1999)Derue, Afonso, Alard, and Albert]{Derue1999}
F.~Derue, C.~Afonso, C.~Alard, and J-N. Albert.
\newblock {Observation of Microlensing towards the Galactic Spiral Arms. EROS
  II 2 year survey}.
\newblock page~11, mar 1999.
\newblock URL \url{http://arxiv.org/abs/astro-ph/9903209}.

\bibitem[Derue et~al.(2002)Derue, Marquette, Lupone, Afonso, Alard, Albert,
  Amadon, Andersen, Ansari, Aubourg, Bareyre, Bauer, Beaulieu, Blanc, Bouquet,
  Char, Charlot, Couchot, Coutures, Ferlet, Fouqu{\'{e}}, Glicenstein, Goldman,
  Gould, Graff, Gros, Ha$\backslash$issinski, Hamilton, Hardin, de~Kat, Kim,
  Lasserre, {Le Guillou}, Lesquoy, Loup, Magneville, Mansoux, Maurice,
  Milsztajn, Moniez, Palanque-Delabrouille, Perdereau, Pr{\'{e}}vot, Regnault,
  Rich, Spiro, Vidal-Madjar, Vigroux, Zylberajch, and {EROS
  Collaboration}]{2002A&A...389..149D}
F~Derue, J.-B. Marquette, S~Lupone, C~Afonso, C~Alard, J.-N. Albert, A~Amadon,
  J~Andersen, R~Ansari, {\'{E}}~Aubourg, P~Bareyre, F~Bauer, J.-P. Beaulieu,
  G~Blanc, A~Bouquet, S~Char, X~Charlot, F~Couchot, C~Coutures, R~Ferlet,
  P~Fouqu{\'{e}}, J.-F. Glicenstein, B~Goldman, A~Gould, D~Graff, M~Gros,
  J~Ha$\backslash$issinski, J.-C. Hamilton, D~Hardin, J~de~Kat, A~Kim,
  T~Lasserre, L~{Le Guillou}, {\'{E}}~Lesquoy, C~Loup, C~Magneville, B~Mansoux,
  {\'{E}}~Maurice, A~Milsztajn, M~Moniez, N~Palanque-Delabrouille, O~Perdereau,
  L~Pr{\'{e}}vot, N~Regnault, J~Rich, M~Spiro, A~Vidal-Madjar, L~Vigroux,
  S~Zylberajch, and {EROS Collaboration}.
\newblock {Observation of periodic variable stars towards the Galactic spiral
  arms by EROS II}.
\newblock \emph{Astronomy and Astrophysics}, 389:\penalty0 149--161, jul 2002.

\bibitem[Faraway et~al.(2014)Faraway, Mahabal, Sun, Wang, Yi, Wang, and
  Zhang]{Faraway2014}
Julian Faraway, Ashish Mahabal, Jiayang Sun, Xiaofeng Wang, Yi, Wang, and
  Lingsong Zhang.
\newblock {Modeling Light Curves for Improved Classification}.
\newblock page~16, jan 2014.
\newblock URL \url{http://arxiv.org/abs/1401.3211}.

\bibitem[Freund et~al.(1999)Freund, Schapire, and Abe]{freund1999short}
Yoav Freund, Robert Schapire, and N~Abe.
\newblock A short introduction to boosting.
\newblock \emph{Journal-Japanese Society For Artificial Intelligence},
  14\penalty0 (771-780):\penalty0 1612, 1999.

\bibitem[Gran et~al.(2015)Gran, Minniti, Saito, Navarrete, D{\'{e}}k{\'{a}}ny,
  McDonald, {Contreras Ramos}, and Catelan]{Gran2015}
F.~Gran, D.~Minniti, R.~K. Saito, C.~Navarrete, I.~D{\'{e}}k{\'{a}}ny,
  I.~McDonald, R.~{Contreras Ramos}, and M.~Catelan.
\newblock {Bulge RR Lyrae stars in the VVV tile b201}.
\newblock \emph{Astronomy {\&} Astrophysics}, 575:\penalty0 A114, mar 2015.
\newblock ISSN 0004-6361.
\newblock \doi{10.1051/0004-6361/201424333}.
\newblock URL \url{http://arxiv.org/abs/1501.00947}.

\bibitem[Huijse et~al.(2012)Huijse, Estevez, Protopapas, Zegers, and
  Principe]{Huijse2012}
Pablo Huijse, Pablo~A. Estevez, Pavlos Protopapas, Pablo Zegers, and
  Jos{\'{e}}~C. Principe.
\newblock {An Information Theoretic Algorithm for Finding Periodicities in
  Stellar Light Curves}.
\newblock \emph{IEEE Transactions on Signal Processing}, 60\penalty0
  (10):\penalty0 5135--5145, oct 2012.
\newblock ISSN 1053-587X.
\newblock \doi{10.1109/TSP.2012.2204260}.
\newblock URL \url{http://arxiv.org/abs/1212.2398}.

\bibitem[Ivezic et~al.(2008)Ivezic, Tyson, Abel, Acosta, Allsman, AlSayyad,
  Anderson, Andrew, Angel, Angeli, Ansari, Antilogus, Arndt, Astier, Aubourg,
  Axelrod, Bard, Barr, Barrau, Bartlett, Bauman, Beaumont, Becker, Becla,
  Beldica, Bellavia, Blanc, Blandford, Bloom, Bogart, Borne, Bosch, Boutigny,
  Brandt, Brown, Bullock, Burchat, Burke, Cagnoli, Calabrese, Chandrasekharan,
  Chesley, Cheu, Chiang, Claver, Connolly, Cook, Cooray, Covey, Cribbs, Cui,
  Cutri, Daubard, Daues, Delgado, Digel, Doherty, Dubois, Dubois-Felsmann,
  Durech, Eracleous, Ferguson, Frank, Freemon, Gangler, Gawiser, Geary, Gee,
  Geha, Gibson, Gilmore, Glanzman, Goodenow, Gressler, Gris, Guyonnet, Hascall,
  Haupt, Hernandez, Hogan, Huang, Huffer, Innes, Jacoby, Jain, Jee, Jernigan,
  Jevremovic, Johns, Jones, Juramy-Gilles, Juric, Kahn, Kalirai, Kallivayalil,
  Kalmbach, Kantor, Kasliwal, Kessler, Kirkby, Knox, Kotov, Krabbendam,
  Krughoff, Kubanek, Kuczewski, Kulkarni, Lambert, Guillou, Levine, Liang, Lim,
  Lintott, Lupton, Mahabal, Marshall, Marshall, May, McKercher, Migliore,
  Miller, Mills, Monet, Moniez, Neill, Nief, Nomerotski, Nordby, O'Connor,
  Oliver, Olivier, Olsen, Ortiz, Owen, Pain, Peterson, Petry, Pierfederici,
  Pietrowicz, Pike, Pinto, Plante, Plate, Price, Prouza, Radeka, Rajagopal,
  Rasmussen, Regnault, Ridgway, Ritz, Rosing, Roucelle, Rumore, Russo, Saha,
  Sassolas, Schalk, Schindler, Schneider, Schumacher, Sebag, Sembroski,
  Seppala, Shipsey, Silvestri, Smith, Smith, Strauss, Stubbs, Sweeney, Szalay,
  Takacs, Thaler, {Van Berg}, Berk, Vetter, Virieux, Xin, Walkowicz, Walter,
  Wang, Warner, Willman, Wittman, Wolff, Wood-Vasey, Yoachim, Zhan, and
  Collaboration]{Ivezic2008}
Z.~Ivezic, J.~A. Tyson, B.~Abel, E.~Acosta, R.~Allsman, Y.~AlSayyad, S.~F.
  Anderson, J.~Andrew, R.~Angel, G.~Angeli, R.~Ansari, P.~Antilogus, K.~T.
  Arndt, P.~Astier, E.~Aubourg, T.~Axelrod, D.~J. Bard, J.~D. Barr, A.~Barrau,
  J.~G. Bartlett, B.~J. Bauman, S.~Beaumont, A.~C. Becker, J.~Becla,
  C.~Beldica, S.~Bellavia, G.~Blanc, R.~D. Blandford, J.~S. Bloom, J.~Bogart,
  K.~Borne, J.~F. Bosch, D.~Boutigny, W.~N. Brandt, M.~E. Brown, J.~S. Bullock,
  P.~Burchat, D.~L. Burke, G.~Cagnoli, D.~Calabrese, S.~Chandrasekharan,
  S.~Chesley, E.~C. Cheu, J.~Chiang, C.~F. Claver, A.~J. Connolly, K.~H. Cook,
  A.~Cooray, K.~R. Covey, C.~Cribbs, W.~Cui, R.~Cutri, G.~Daubard, G.~Daues,
  F.~Delgado, S.~Digel, P.~Doherty, R.~Dubois, G.~P. Dubois-Felsmann,
  J.~Durech, M.~Eracleous, H.~Ferguson, J.~Frank, M.~Freemon, E.~Gangler,
  E.~Gawiser, J.~C. Geary, P.~Gee, M.~Geha, R.~R. Gibson, D.~K. Gilmore,
  T.~Glanzman, I.~Goodenow, W.~J. Gressler, P.~Gris, A.~Guyonnet, P.~A.
  Hascall, J.~Haupt, F.~Hernandez, C.~Hogan, D.~Huang, M.~E. Huffer, W.~R.
  Innes, S.~H. Jacoby, B.~Jain, J.~Jee, J.~G. Jernigan, D.~Jevremovic,
  K.~Johns, R.~L. Jones, C.~Juramy-Gilles, M.~Juric, S.~M. Kahn, J.~S. Kalirai,
  N.~Kallivayalil, B.~Kalmbach, J.~P. Kantor, M.~M. Kasliwal, R.~Kessler,
  D.~Kirkby, L.~Knox, I.~Kotov, V.~L. Krabbendam, S.~Krughoff, P.~Kubanek,
  J.~Kuczewski, S.~Kulkarni, R.~Lambert, L.~Le Guillou, D.~Levine, M.~Liang,
  K-T. Lim, C.~Lintott, R.~H. Lupton, A.~Mahabal, P.~Marshall, S.~Marshall,
  M.~May, R.~McKercher, M.~Migliore, M.~Miller, D.~J. Mills, D.~G. Monet,
  M.~Moniez, D.~R. Neill, J-Y. Nief, A.~Nomerotski, M.~Nordby, P.~O'Connor,
  J.~Oliver, S.~S. Olivier, K.~Olsen, S.~Ortiz, R.~E. Owen, R.~Pain, J.~R.
  Peterson, C.~E. Petry, F.~Pierfederici, S.~Pietrowicz, R.~Pike, P.~A. Pinto,
  R.~Plante, S.~Plate, P.~A. Price, M.~Prouza, V.~Radeka, J.~Rajagopal,
  A.~Rasmussen, N.~Regnault, S.~T. Ridgway, S.~Ritz, W.~Rosing, C.~Roucelle,
  M.~R. Rumore, S.~Russo, A.~Saha, B.~Sassolas, T.~L. Schalk, R.~H. Schindler,
  D.~P. Schneider, G.~Schumacher, J.~Sebag, G.~H. Sembroski, L.~G. Seppala,
  I.~Shipsey, N.~Silvestri, J.~A. Smith, R.~C. Smith, M.~A. Strauss, C.~W.
  Stubbs, D.~Sweeney, A.~Szalay, P.~Takacs, J.~J. Thaler, R.~{Van Berg},
  D.~Vanden Berk, K.~Vetter, F.~Virieux, B.~Xin, L.~Walkowicz, C.~W. Walter,
  D.~L. Wang, M.~Warner, B.~Willman, D.~Wittman, S.~C. Wolff, W.~M. Wood-Vasey,
  P.~Yoachim, H.~Zhan, and for the~LSST Collaboration.
\newblock {LSST: from Science Drivers to Reference Design and Anticipated Data
  Products}.
\newblock page~39, may 2008.
\newblock URL \url{http://arxiv.org/abs/0805.2366}.

\bibitem[Jordan and Jacobs(1993)]{Jordan1993}
Michael~I Jordan and Robert~A Jacobs.
\newblock {Hierarchical Mixtures of Experts and the EM Algorithm}.
\newblock \penalty0 (1440), 1993.

\bibitem[Jordan and Jacobs(1994)]{jordan1994hierarchical}
Michael~I Jordan and Robert~A Jacobs.
\newblock Hierarchical mixtures of experts and the em algorithm.
\newblock \emph{Neural computation}, 6\penalty0 (2):\penalty0 181--214, 1994.

\bibitem[Kaiser(2004)]{Kaiser2004SPIE}
N~Kaiser.
\newblock {Pan-STARRS: a wide-field optical survey telescope array}.
\newblock In {J.{\~{}}M.{\~{}}Oschmann Jr.}, editor, \emph{Society of
  Photo-Optical Instrumentation Engineers (SPIE) Conference Series}, volume
  5489 of \emph{Society of Photo-Optical Instrumentation Engineers (SPIE)
  Conference Series}, pages 11--22, oct 2004.
\newblock \doi{10.1117/12.552472}.

\bibitem[Kelly et~al.(2009)Kelly, Bechtold, and Siemiginowska]{Kelly:2009}
B.{\~{}}C. Kelly, J~Bechtold, and A~Siemiginowska.
\newblock {Are the Variations in Quasar Optical Flux Driven by Thermal
  Fluctuations?}
\newblock 698:\penalty0 895--910, 2009.

\bibitem[Kim et~al.(2011)Kim, Protopapas, Byun, Alcock, Khardon, and
  Trichas]{Kim:2011}
D.-W. Kim, P~Protopapas, Y.-I. Byun, C~Alcock, R~Khardon, and M~Trichas.
\newblock {Quasi-stellar Object Selection Algorithm Using Time Variability and
  Machine Learning: Selection of 1620 Quasi-stellar Object Candidates from
  MACHO Large Magellanic Cloud Database}.
\newblock \emph{The Astrophysical Journal}, 735, 2011.

\bibitem[Kim et~al.(2012)Kim, Protopapas, Trichas, Rowan-Robinson, Khardon,
  Alcock, and Byun]{Kim:2012}
D.-W. Kim, P~Protopapas, M~Trichas, M~Rowan-Robinson, R~Khardon, C~Alcock, and
  Y.-I. Byun.
\newblock {A Refined QSO Selection Method Using Diagnostics Tests: 663 QSO
  Candidates in the Large Magellanic Cloud}.
\newblock \emph{The Astrophysical Journal}, 747, 2012.

\bibitem[Kim et~al.(2014)Kim, Protopapas, Bailer-Jones, Byun, Chang, Marquette,
  and Shin]{Kim2014}
Dae-Won Kim, Pavlos Protopapas, Coryn A.~L. Bailer-Jones, Yong-Ik Byun, Seo-Won
  Chang, Jean-Baptiste Marquette, and Min-Su Shin.
\newblock {The EPOCH Project}.
\newblock \emph{Astronomy {\&} Astrophysics}, 566:\penalty0 A43, jun 2014.
\newblock ISSN 0004-6361.
\newblock \doi{10.1051/0004-6361/201323252}.
\newblock URL \url{http://arxiv.org/abs/1403.6131}.

\bibitem[Kuncheva(2007)]{Kuncheva2007}
L.~I. Kuncheva.
\newblock {Combining Pattern Classifiers: Methods and Algorithms (Kuncheva,
  L.I.; 2004) [book review]}.
\newblock \emph{IEEE Transactions on Neural Networks}, 18\penalty0 (3), 2007.
\newblock ISSN 1045-9227.
\newblock \doi{10.1109/TNN.2007.897478}.

\bibitem[Kuncheva(2004)]{kuncheva2004combining}
Ludmila~I Kuncheva.
\newblock \emph{Combining pattern classifiers: methods and algorithms}.
\newblock John Wiley \& Sons, 2004.

\bibitem[Kuncheva and Whitaker(2003)]{Kuncheva_Whitaker}
Ludmila~I. Kuncheva and Christopher~J. Whitaker.
\newblock {Measures of diversity in classifier ensembles and their relationship
  with the ensemble accuracy}.
\newblock \emph{Machine learning}, 51\penalty0 (2):\penalty0 181--207, 2003.
\newblock ISSN 0885-6125.
\newblock URL \url{http://cat.inist.fr/?aModele=afficheN{\&}cpsidt=14668817}.

\bibitem[LeBlanc and Tibshirani(1996)]{leblanc1996combining}
Michael LeBlanc and Robert Tibshirani.
\newblock Combining estimates in regression and classification.
\newblock \emph{Journal of the American Statistical Association}, 91\penalty0
  (436):\penalty0 1641--1650, 1996.

\bibitem[Lin and Liu(2005)]{lin2005robust}
Yen-Yu Lin and Tyng-Luh Liu.
\newblock Robust face detection with multi-class boosting.
\newblock In \emph{Computer Vision and Pattern Recognition, 2005. CVPR 2005.
  IEEE Computer Society Conference on}, volume~1, pages 680--687. IEEE, 2005.

\bibitem[Long et~al.(2012)Long, Karoui, Rice, Richards, and Bloom]{Long2012}
James~P. Long, Noureddine~El Karoui, John~A. Rice, Joseph~W. Richards, and
  Joshua~S. Bloom.
\newblock {Optimizing Automated Classification of Variable Stars in New
  Synoptic Surveys}.
\newblock \emph{Publications of the Astronomical Society of the Pacific},
  124\penalty0 (913):\penalty0 280--295, mar 2012.
\newblock ISSN 00046280.
\newblock \doi{10.1086/664960}.
\newblock URL \url{http://arxiv.org/abs/1201.4863}.

\bibitem[Meir(1996)]{Meir1996}
Ronny Meir.
\newblock {Non-linear Models for Time Series Using Mixtures of Experts 1
  Introduction}.
\newblock \penalty0 (May), 1996.

\bibitem[Nun et~al.(2014)Nun, Pichara, Protopapas, and Kim]{Nun2014}
Isadora Nun, Karim Pichara, Pavlos Protopapas, and Dae-Won Kim.
\newblock {SUPERVISED DETECTION OF ANOMALOUS LIGHT CURVES IN MASSIVE
  ASTRONOMICAL CATALOGS}.
\newblock \emph{The Astrophysical Journal}, 793\penalty0 (1):\penalty0 23, sep
  2014.
\newblock ISSN 1538-4357.
\newblock \doi{10.1088/0004-637X/793/1/23}.
\newblock URL \url{http://arxiv.org/abs/1404.4888}.

\bibitem[Nun et~al.(2015)Nun, Protopapas, Sim, Zhu, Dave, Castro, and
  Pichara]{Nun2015}
Isadora Nun, Pavlos Protopapas, Brandon Sim, Ming Zhu, Rahul Dave, Nicolas
  Castro, and Karim Pichara.
\newblock {FATS: Feature Analysis for Time Series}.
\newblock may 2015.
\newblock URL \url{http://arxiv.org/abs/1506.00010}.

\bibitem[Pichara and Protopapas(2013)]{Pichara_Miss:2013}
K.~Pichara and P.~Protopapas.
\newblock {Automatic Classification of Variable Stars in Catalogs with Missing
  Data}.
\newblock \emph{The Astrophysical Journal}, 777:\penalty0 83, 2013.

\bibitem[Pichara et~al.(2012)Pichara, Protopapas, Kim, Marquette, and
  Tisserand]{Pichara_QSO:2012}
K~Pichara, P~Protopapas, D~Kim, J~Marquette, and P~Tisserand.
\newblock {An improved quasar detection method in EROS-2 and MACHO LMC
  datasets}.
\newblock \emph{Monthly Notices of the Royal Academy Society}, 18\penalty0
  (September):\penalty0 1--18, 2012.

\bibitem[Protopapas et~al.(2015)Protopapas, Huijse, Est{\'{e}}vez, Zegers,
  Pr{\'{\i}}ncipe, and Marquette]{Protopapas2015}
Pavlos Protopapas, Pablo Huijse, Pablo~A. Est{\'{e}}vez, Pablo Zegers,
  Jos{\'{e}}~C. Pr{\'{\i}}ncipe, and Jean-Baptiste Marquette.
\newblock {A NOVEL, FULLY AUTOMATED PIPELINE FOR PERIOD ESTIMATION IN THE EROS
  2 DATA SET}.
\newblock \emph{The Astrophysical Journal Supplement Series}, 216\penalty0
  (2):\penalty0 25, jan 2015.
\newblock ISSN 1538-4365.
\newblock \doi{10.1088/0067-0049/216/2/25}.
\newblock URL \url{http://arxiv.org/abs/1412.1840}.

\bibitem[Quinlan(1993)]{Quinlan:1993}
J~Quinlan.
\newblock \emph{{C4.5: programs for machine learning}}.
\newblock Morgan Kaufmann Publishers Inc., 1993.

\bibitem[Quinlan(1986)]{Quinlan:1986}
R~Quinlan.
\newblock {Induction of Decision Trees}.
\newblock \emph{Machine Learning}, 1\penalty0 (1):\penalty0 81--106, 1986.

\bibitem[Rasmussen and Ghahramani(1991)]{Rasmussen1991}
Carl~Edward Rasmussen and Zoubin Ghahramani.
\newblock {Infinite Mixtures of Gaussian Process Experts}.
\newblock 1991.

\bibitem[Richards et~al.(2011)Richards, Starr, Butler, Bloom, Brewer,
  Crellin-Quick, Higgins, Kennedy, and Rischard]{Richards:2011}
J.{\~{}}W. Richards, D.{\~{}}L. Starr, N.{\~{}}R. Butler, J.{\~{}}S. Bloom,
  J.{\~{}}M. Brewer, A~Crellin-Quick, J~Higgins, R~Kennedy, and M~Rischard.
\newblock {On Machine-learned Classification of Variable Stars with Sparse and
  Noisy Time-series Data}.
\newblock \emph{The Astrophysical Journal}, 733, 2011.

\bibitem[Tisserand et~al.(2007)Tisserand, {Le Guillou}, Afonso, Albert,
  Andersen, Ansari, Aubourg, Bareyre, Beaulieu, Charlot, Coutures, Ferlet,
  Fouqu{\'{e}}, Glicenstein, Goldman, Gould, Graff, Gros, Haissinski,
  Hamadache, de~Kat, Lasserre, Lesquoy, Loup, Magneville, Marquette, Maurice,
  Maury, Milsztajn, Moniez, Palanque-Delabrouille, Perdereau, Rahal, Rich,
  Spiro, Vidal-Madjar, Vigroux, Zylberajch, and {EROS-2
  Collaboration}]{Tisserand:2007}
P~Tisserand, L~{Le Guillou}, C~Afonso, J.{\~{}}N. Albert, J~Andersen, R~Ansari,
  {\'{E}}~Aubourg, P~Bareyre, J.{\~{}}P. Beaulieu, X~Charlot, C~Coutures,
  R~Ferlet, P~Fouqu{\'{e}}, J.{\~{}}F. Glicenstein, B~Goldman, A~Gould,
  D~Graff, M~Gros, J~Haissinski, C~Hamadache, J~de~Kat, T~Lasserre,
  {\'{E}}~Lesquoy, C~Loup, C~Magneville, J.{\~{}}B. Marquette, {\'{E}}~Maurice,
  A~Maury, A~Milsztajn, M~Moniez, N~Palanque-Delabrouille, O~Perdereau,
  Y.{\~{}}R. Rahal, J~Rich, M~Spiro, A~Vidal-Madjar, L~Vigroux, S~Zylberajch,
  and {EROS-2 Collaboration}.
\newblock {Limits on the Macho content of the Galactic Halo from the EROS-2
  Survey of the Magellanic Clouds}.
\newblock \emph{Astronomy and Astrophysics}, 469:\penalty0 387--404, 2007.

\bibitem[Udalski et~al.(2008)Udalski, Szymanski, Soszynski, and
  Poleski]{Udalski2008AcA}
A~Udalski, M.{\~{}}K. Szymanski, I~Soszynski, and R~Poleski.
\newblock {The Optical Gravitational Lensing Experiment. Final Reductions of
  the OGLE-III Data}.
\newblock \emph{Acta Astronomica}, 58:\penalty0 69--87, jun 2008.

\bibitem[Wolpert(1992)]{wolpert1992stacked}
David~H Wolpert.
\newblock Stacked generalization.
\newblock \emph{Neural networks}, 5\penalty0 (2):\penalty0 241--259, 1992.

\bibitem[Xu et~al.(1995)Xu, Jordan, and Hinton]{xu1995alternative}
Lei Xu, Michael~I Jordan, and Geoffrey~E Hinton.
\newblock An alternative model for mixtures of experts.
\newblock \emph{Advances in neural information processing systems}, pages
  633--640, 1995.

\bibitem[Zehnder et~al.(2008)Zehnder, Koller-Meier, and
  Van~Gool]{zehnder2008efficient}
Philipp Zehnder, Esther Koller-Meier, and Luc~J Van~Gool.
\newblock An efficient shared multi-class detection cascade.
\newblock In \emph{BMVC}, pages 1--10, 2008.

\end{thebibliography}

\end{document}